\documentclass[12pt]{article}
\usepackage{cite}
\usepackage[dvipdfm]{graphicx}
\usepackage{amssymb}
\usepackage{amsmath}
\usepackage{epsfig}
\usepackage{dcolumn}
\usepackage{rotating}
\usepackage{color}
\definecolor{rosso}{cmyk}{0,1,1,0.4}
\definecolor{rossos}{cmyk}{0,1,1,0.55}
\definecolor{rossoc}{cmyk}{0,1,1,0.2}
\definecolor{blu}{cmyk}{1,1,0,0.3}
\definecolor{blus}{cmyk}{1,1,0,0.6}
\definecolor{bluc}{cmyk}{1,1,0,0.1}
\definecolor{verde}{cmyk}{0.92,0,0.59,0.25}
\definecolor{verdec}{cmyk}{0.92,0,0.59,0.15}
\definecolor{verdes}{cmyk}{0.92,0,0.59,0.7}

\newcommand{\ba}{\begin{eqnarray}}
\newcommand{\ea}{\end{eqnarray}}
\newcommand{\be}{\begin{equation}}
\newcommand{\ee}{\end{equation}}
\newcommand{\bi}{\begin{itemize}}
\newcommand{\ei}{\end{itemize}}
\newcommand{\al}{\alpha}
\newcommand{\bt}{\beta}
\newcommand{\ga}{\gamma}

\newcommand{\la}{\lambda}
\newcommand{\ka}{\kappa}

\newcommand{\sa}{\sigma}
\newcommand{\en}{\epsilon}

\newcommand{\Ga}{\Gamma}

\newcommand{\Da}{\Delta}
\newcommand{\Oa}{\Omega}
\newcommand{\La}{\Lambda}

\newcommand{\cR}{{\cal R}}
\newcommand{\cO}{{\cal O}}

\newcommand{\cT}{{\cal T}}

\newcommand{\ra}{\rightarrow}

\newcommand{\LF}{\left(}
\newcommand{\RF}{\right)}
\newcommand{\LT}{\left[}
\newcommand{\RT}{\right]}
\newcommand{\Ld}{\left.}
\newcommand{\Rd}{\right.}

\newcommand{\nb}{\bar{n}}

\newcommand{\3}{\frac{1}{3}}
\newcommand{\4}{\frac{1}{4}}
\newcommand{\6}{\frac{1}{6}}

\newcommand{\mx}{\mbox}
\newcommand{\mt}{\mathtt}
\newcommand{\mand}{\mx{ and }}

\newcommand{\where}{\mx{ where }}
\newcommand{\with}{\mx{ with }}

\newcommand{\tot}{\mt{tot}}
\newcommand{\ie}{{\it i.e.\ }}

\newcommand{\vs}{\vspace{5mm}\\}
\newcommand{\non}{\nonumber\\}

\newcommand{\rhob}{\rho_{\bullet}}
\newcommand{\xeq}{x_{\mt{min}}}

\newcommand{\bb}{\bar{\bt}}
\begin{document}
\tolerance=100000
\thispagestyle{empty}
\vspace{1cm}

\begin{center}
{\LARGE \bf
Numerically Investigating the Emergent Cyclic Inflation Scenario
 \\ [0.15cm]
}
\vskip 2cm
{\large William Duhe\footnote{wduhe@loyno.edu}} \&
{\large Tirthabir Biswas\footnote{tbiswas@loyno.edu}},

\vskip 7mm
{\it Department of Physics, Loyola University, \\ 6300 St. Charles Avenue\\
New Orleans, LA, 70118, USA}
\vskip 3mm
\vskip 3mm
\end{center}
\date{\today}

\begin{abstract}
We provide a comprehensive numerical study of the Emergent Cyclic Inflation scenario. This is a scenario where instead of traditional monotonic slow roll inflation, the universe expands over numerous short asymmetric cycles due to the production of entropy via interactions among different species. This is one of the very few scenarios of inflation which provides a nonsingular geodesically complete space-time and does not require any ``reheating'' mechanism.
\end{abstract}

\newpage

\setcounter{page}{1}

\tableofcontents
\section{Introduction}
The inflationary paradigm has been one of the cornerstones of  success of the Standard cosmological model, for recent reviews see~\cite{infl-recent,infl-rev}. However, the inflationary paradigm is incomplete in more than one way: While the slowly varying scalar potential energy driven inflation have been the favorite model, there have been a plethora of other mechanisms that has been considered in the literature, here are but a few examples~\cite{Garriga,warm,padic,vector-infl,fading-gravity,fast-roll}. Even within the general class of scalar field driven slow-roll inflation, models abide in plenty~\cite{infl-recent}. While the observational data has been narrowing down the list of viable inflationary models, and this situation should dramatically improve with more data on the gravity wave spectrum, it is fair to say that we still don't have any obvious winner among all the inflationary mechanisms/models. One can blame this lack of closure to our ignorance of the ``fundamental'' particle/string theory which makes a top-down approach to inflationary model building challenging, but the inflationary space-time itself is also ``geodesically incomplete''. Essentially, if one tries to track a particle trajectory into the past, the geodesic ends abruptly at a finite proper time; and this  has been known since early days of inflationary cosmology~\footnote{For other well known conceptual issues related to inflation see~\cite{robert-criticism,neil-criticism}.}~\cite{Borde,Linde}. It is clear that there had to have been a ``pre-inflationary'' phase.

Naively it may seem that since the inflationary regime is an ``attractor'', the pre-inflationary phase can simply be ignored without affecting any of the advantages of inflation. Indeed, if one takes the view that the pre-inflationary phase was a monotonic expansion culminating in the past (possibly) in a quantum space-time regime, which we are yet to understand fully, the arguments in favor of inflation seem robust. However, as argued in~\cite{essay}, if instead one believes that an ``effective space-time'' description remains valid even at the Planckian era, and that the quantum gravity/stringy corrections are able to resolve the Big Bang singularity via the ``Big Bounce''~\footnote{For quantum gravity/string theory inspired models which exhibit this property see for instance~\cite{loop,ashtekar,Shtanov,BMS,BGKM,Cai-bounce,Cai-saridakis}.} so that our phase of expansion is preceded by a phase of contraction, several advantages of the inflationary paradigm suddenly become questionable. For instance, if the Big Bounce is preceded by a long phase of contraction, the rapid growth of anisotropies as $a^{-6}$ makes it very hard to imagine how the universe could  ever have reached an isotropic patch  where inflation could become activated; typically long before an FLRW bounce can be reached the universe would become completely anisotropic entering into a chaotic Mixmaster phase~\cite{mixmaster,misner-thorne-wheeler}.

In this scenario it  also becomes unclear why one should use the Bunch-Davis vacuum initial conditions for the calculation of  the primordial spectrum of fluctuations seeding the  CMBR anisotropies. These calculations agree remarkably well  with the observations~\cite{WMAP,Planck}, but if one has a prior contracting phase, one ought to start with initial seed fluctuations in that contracting phase, track it through the bounce, and then along the inflationary expansion; there are no general arguments known to the authors which demonstrate that the resulting spectrum in such a case will still be nearly scale-invariant as observed in CMBR. In fact, cosmologists have been looking at non-inflationary mechanisms to generate the desired near scale-invariant spectrum utilizing the contraction/bounce phase, notable examples being the ekpyrotic~\cite{ekpyrotic}, matter-bounce~\cite{matter-bounce} and Hagedorn-bounce~\cite{BKM-hagedorn,BBMS} scenarios.

In the presence of a bounce mechanism, to avoid these theoretical challenges a radically different inflationary mechanism was  proposed in a series of articles~\cite{emergent-cyclic,AB-cyclic,cyclic-inflation,cyclic-prediction,BKM-exit,BKM-cyclic}  where one hopes to realize a geodesically complete non-singular inflationary paradigm via a series of asymmetric short cycles. Here is the Emergent-Cyclic-Inflation (ECI) paradigm  in brief:
\begin{itemize}
\item The universe ``begins'' in a quasi-periodic phase of oscillation around a finite sized universe. These oscillations  continue all the way to past infinity becoming more and more symmetric and periodic~\cite{emergent-cyclic}. This is very similar to the ``emergent universe'' scenario advocated in~\cite{emergent1,emergent2}. The potential energy is negative but it remains subdominant as compared to the negative curvature density associated with a closed universe which cancels the matter densities to provide the turnarounds.
\item This is followed by a phase of asymmetric cyclic growth when the negative potential energy takes over from the negative curvature density in providing the turnarounds. Over many cycles the space-time resembles the inflationary growth~\cite{cyclic-inflation}. In every cycle (same time period) the volume of the universe grows by the same factor. This occurs because the entropy in each of these cycles due to interaction between different particle species increases by the same factor:
     \be
     {S_{n+1}\over S_n}=1+3\ka\ ,
     \ee
     where $S_n$ denotes the entropy of the ``$n$th'' cycle. Since the total entropy is proportional to the volume, this means that for small $\ka$, the scale-factor (at the same temperature) in consecutive cycles increases by a factor $\ka$.
     During the cyclic inflation (CI) phase the magnitude of the curvature density is small as compared to that of the cosmological constant.
\item The CI phase ``graceful exits'' via scalar field dynamics which can classically propel the universe from a negative to a positive potential energy region, given the right ``condition''~\cite{BKM-exit} .
\item Henceforth a monotonic phase of expansion begins as in the Standard Cosmological model. No reheating is required in the ECI paradigm as radiation remains the dominant energy density component throughout the entire evolution.
\end{itemize}
The ECI paradigm provides a nonsingular geodesically complete inflationary cosmology, does not require any reheating (which has it's own challenges~\cite{infl-rev}), and interestingly predicts distinctive signatures in the form of small oscillatory wiggles in the power spectrum~\cite{cyclic-prediction}~\footnote{There have been other cyclic models~\cite{cyclic-multiverse1,cyclic-multiverse2} which also looked at the flow of perturbations through various cycles.}. More recently, the BICEP data~\cite{bicep} has provided an additional motivation to consider bouncing/cyclic models as often in these scenarios one can have large scalar-to-tensor ratios (see for instance~\cite{gw-cyclic,BKM-hagedorn}) and can also naturally produce a blue gravity wave spectrum~\footnote{The gravity wave spectrum is proportional to the absolute value of $H$. During inflation, $H$ is slowly decreasing giving rise to a red spectrum, while during contraction, $|H|$ increases yielding a blue spectrum.} which is mildly favored at present~\cite{blue-gw1,blue-gw2,blue-gw3}. This has been  discussed in the specific context of cyclic inflation model in~\cite{BKM-hagedorn} which suggests that extensions of the CI model will be worth exploring in the future. Finally, the CI mechanism also provides new avenues of constructing phenomenologically viable inflationary models where the initial potential energy could be negative. With the possible exception of the monopole problem the ECI can address all the old cosmological puzzles as in standard inflation.

While the cosmology described above was proposed and elaborated on in a series of papers~\cite{emergent-cyclic,AB-cyclic,cyclic-inflation,cyclic-prediction,BKM-exit,BKM-cyclic}, several arguments were either qualitative in nature or relied on naive estimates. Our aim in this paper is to investigate the background cosmology and test the viability of the ECI scenario comprehensively using numerical methods. In the process we also hope clarify the various parameter dependencies on observable features in the CMBR spectrum, such as the amplitude and size of the oscillatory wiggles~\cite{cyclic-prediction}. This would be useful when we try to fit the model to the recently released Planck data using Monte Carlo simulations. Also, this is the first time when we  examine whether the ``emergence'' mechanism proposed in~\cite{emergent-cyclic} meshes consistently with the cyclic inflation set-up~\footnote{The scenario we will present here is  similar to what was considered in~\cite{emergent-cyclic}, except that in ECI one has a  radiation dominated universe, where as in the set-up discussed in~\cite{emergent-cyclic} the dominant contribution to energy density  came from the NR species. The importance of having a radiation dominated universe is that it's pressure is able to avoid the Black-hole over-production problem, recurrent in cyclic cosmologies,  see~\cite{cyclic-inflation} for a detailed argument.}. Finally, we provide a general framework to study interacting thermal fluids in contracting backgrounds which may be useful in other cyclic/bouncing universe models which continues to be the most popular alternative to the inflationary paradigm. For instance, through the course of our numerical investigation we found that the naive estimate of the temperature at which thermal equilibrium can be restored in a contracting universe is a little lower than what our numerical simulations show. On the other hand, once such an equilibrium is established, the various  components (relativistic by now) continue to track the equilibrium densities all the way up to Planckian bounce densities, even though it is well known that they should not be able to maintain thermal equilibrium at such high temperatures~\cite{kolb-turner}. This is essentially the  analogue of the phenomena that is observed in the expanding phase involving the CMBR photons; the fact that they still obey Planck's blackbody distribution even though they have been traveling essentially freely for the last 14 billion years or so. This finding may be relevant for different cosmological scenarios: For instance, one of the criticisms~\cite{hagedorn-interaction} of the String Gas Cosmology framework~\cite{sgc-review} is that it is very difficult to establish thermal equilibrium among all the stringy excitations in the Hagedorn phase. Our analysis suggests that there may be a way around this problem if one includes a prior contracting phase.

Our paper is organized as follows: In section~\ref{sec:equilibrium}, we will first introduce a toy model which can realize the CI mechanism and test whether after each turnaround it is possible to recreate  the nonrelativistic particles from the massless degrees of freedom in the contracting phase, and whether thermal equilibrium can be re-established. This is key in being able to sustain the inflationary growth over many cycles. In section~\ref{sec:growth}, we will look at the coupled dynamics of the massless and massive species along with the scale factor in a given cycle, and track the growth of the scale factor due to entropy production. Next, in section~\ref{sec:ECI}, we will  verify whether the universe can indeed sustain the inflationary growth over numerous cycles.  We will also look at the merging of the cyclic inflation scenario with the emergence mechanism~\cite{emergent-cyclic}. We will conclude with a perspective on future research directions in section~\ref{sec:conclusions}.
\section{Recycling the Universe}\label{sec:equilibrium}
\subsection{ECI, Motivation \& Overview}
As argued in the introduction (also see~\cite{essay}), if the universe has a built-in mechanism to bounce from a phase of contraction to a phase of expansion, some of the advantages of the inflationary paradigm are lost (or at least need to be revisited). One way of salvaging the scenario is by relaxing the pre-condition that our universe ``started'' with positive potential energy. Actually, there is yet no known ``theoretical'' reasons as to why the universe should have positive potential energy. String theory, the leading candidate for a consistent theory of quantum gravity, naturally predicts the existence of negative energy vacua, and in fact, it has been quite a challenge to find ways that may lead to positive vacuum energies in the string theory framework~\cite{Kachru:2003sx}. The same is also true for supergravity theories~\cite{sugra-ds}.
In general, if one looks at the potential energy coming from all the moduli in any fundamental theory,  one would expect to have both negative and positive potential regions, and local minima's dispersed liberally~\cite{flux-compactification}. What we do know however from cosmological observations, is that currently the universe has a positive cosmological constant~\cite{Perlmutter:1997zf}. We will see later how the ECI paradigm can reconcile this apparent contradiction between the early and late time cosmology.

Presently, let us try to investigate the dynamics of a ``small'' but relatively smooth patch in the ``Early Universe''with  negative vacuum energy density, $\La<0$. Typically, with such a patch apart from the vacuum/potential energy, $\La$, one expects to associate energy density components such as spatial curvature, $\rho_k\sim a^{-2}$,  massless degrees of freedom (radiation), $\rho_r\sim a^{-4}$,  perhaps some massive modes as well, $\rho_m\sim a^{-3}$, and possibly small amounts of anisotropies~\footnote{One could also add energy coming from kinetic energy of massless scalar fields, but it behaves in a manner very similar to that of the anisotropies, $\rho_{\phi}\sim a^{-6}$, and therefore can be clubbed with $\rho_a$.}, $\rho_a\sim a^{-6}$. If $|\La|> \rho_r+\rho_m+\rho_k$, FRW evolution is inconsistent, most likely the universe would be stuck in a static anti-de Sitter like universe containing massless and massive excitations. Let us then look at the opposite case when $|\La|< \rho_r+\rho_m+\rho_k$. In this case, as the universe expands, all the matter components would dilute and eventually cancel the negative cosmological constant causing the universe to turnaround and start contracting. As the energy density increases to Planckian densities, according to our prior assumption, quantum gravity effects would prevent a Big Crunch singularity and usher in the next phase of expansion. This story will keep on repeating leading to a cyclic model.

At first glance, such a cyclic universe is completely inconsistent with our universe. If $|\La|$ is large, given possibly by the string/GUT scale, typically a few orders of magnitude below the Planck scale, then each of the cycles would only last a very short time $\tau\sim M_p/\sqrt{\La}\sim 10^{-33}$ s, much too short for any realistic cosmology. Making $|\La|$ small to allow for structure formation doesn't help either because it will be in violent conflict with the current cosmic speed-up data. What turns out to be a natural savior is  in any realistic high energy physics model one expects interactions between different particle species which  tends to create entropy. Entropy can only increase monotonically according to the second law of thermodynamics, this then provides a simple way of breaking the periodicity of the evolution. Actually, this was precisely what Tolman pointed out in the 1930's giving rise to Tolman's entropy problem for cyclic models~\cite{tolman1,tolman2}, but we can now  use the entropy production to our advantage. As was first pointed out in~\cite{cyclic-inflation}, and will be further elaborated on in this paper, entropy tends to increase by the same factor in every cycle, while the time period of the cycles remains a constant since it is governed by $\La$ (which for simplicity we will assume to be a constant). This means that the universe must be growing by the same factor in every cycle giving rise to an overall inflationary growth! Most of the advantages of the standard inflationary paradigm rolls over to this scenario,  including the production of near scale-invariant density fluctuations, see~\cite{cyclic-inflation,cyclic-prediction,essay} for details.

As one must, with any inflationary mechanism, we need to provide a ``graceful exit''. In  an expanding background it is known that energy densities of scalar fields can only decrease, and hence once the universe is in a negative energy phase, there is no way for it to claw back up to the positive region. However, in contracting phases the reverse is true {\it i.e.},  the  energy density  increases, and in~\cite{BKM-exit} (see also~\cite{Felder,Mulryne:2005ef,Lidsey:2006md,Nunes:2005ra,Piao:2004me}) it was demonstrated that  contracting phases can indeed facilitate a transition from negative to positive potential energy regions. Thus our universe could have been inflating while ``exploring'' the negative potential regions when at some point  it made a classical jump to a positive potential energy region, in the process providing a graceful exit from the CI phase and  ushering in an everlasting monotonic phase of expansion thereafter!

There is a final twist to  the  beginning of the story. Let us revisit the issue of geodesic completeness in the context of the CI scenario. If one tracks, say, the maximum of the oscillating space time, then one finds that it has the traditional inflationary trajectory. The problem of past geodesic incompleteness comes back to haunt us! Fortunately, there may be a natural resolution for a closed universe~\footnote{The model only works if the universe is closed. For an open patch, one ends up with a very different dynamics and an universe nothing in common to ours.}. In this case,  as one goes back in cycles there comes a point when the negative curvature energy density becomes more important than the negative vacuum energy density. (Curvature density blue shifts as $a^{-2}$, while the vacuum energy density remains a constant.) Once this happens, the universe no longer turns around due to the negative vacuum energy density, but before, when $\rho_r+\rho_k=0$. The temperature of the turnaround increases making the universe spend less and less time in the out-of-equilibrium phases (massive particles only fall out-of-equilibrium when their temperature falls below their masses). This ensures that the entropy increase becomes less and less as we go  further back  in cycles, and eventually the universe asymptotes to a periodic evolution as $t\ra -\infty$. The space-time is, in fact, very reminiscent of the emergent universe scenario advocated in~\cite{emergent1,emergent2}, and this is reason why refer to our cosmological model  as ``emergent-cyclic-inflation'' or ECI.

\subsection{A prototype model}
To illustrate the basics of the CI mechanism it is sufficient to consider a universe  where we have a negative cosmological constant~\footnote{The evolution of $\La$ as a function of the various scalar fields that may be present in a fundamental theory is only relevant in the discussion of the spectral tilt, as $\La$ controls the amplitude of the primordial spectrum, and in the discussion of the graceful exit mechanism from the CI phase. These issues have been addressed in~\cite{cyclic-prediction} and~\cite{BKM-exit}, but we plan on performing a comprehensive numerical analysis that will be required when we fit our model to the Planck data.} ($-\La$), and the ``matter content'' of the universe  consists of a single non-relativistic (NR) species, $\psi$,  along with relativistic degrees of freedom, collectively denoted by $X$. To obtain the cyclic evolution, we will assume that during contraction when a critical Planckian energy density, $\rhob$,  is reached, the universe bounces back nonsingularly to a phase of expansion. While the Big Bounce is still very much a conjecture, and there is yet to be a completely convincing mechanism in place, several different bouncing mechanism  have been considered in the literature, and in many of these models a bounce occurs when the energy density reaches close to the Planck density, see for instance~\cite{BMS,BKM,loop,ashtekar,Freese,Baum,Shtanov,palatini-bounce}. For the purpose of our paper we will simply assume the existence of such a mechanism. Now, the success of the CI mechanism is based on being able to increase the entropy on the universe by a constant factor in each cycle. This entropy production is achieved via  energy exchange between two components of matter which are not in equilibrium with each other. In our specific toy model this process will  correspond to simple decays of some massive particles to massless degrees of freedom at low temperatures (compared to Planckian bounce temperature) when the nonrelativistic massive species is not in thermal equilibrium with the ambient radiative fluid.  Therefore, the precise nature of the bouncing cosmology will have very little effect on the  particle physics that is relevant for entropy production driving CI. Given our ignorance about the bouncing mechanism this is clearly an advantage of the CI scenario.

Away from the bounce when GR is valid, the Hubble equation corresponding to flat  Friedman-Lemaitre-Robertson-Walker (FLRW) cosmology reads
\be
H^2={\rho\over 3M_p^2}\ ,
\label{hubble}
\ee
where $\rho$ is the total energy density  given by
\be
\rho=\rho_m+\rho_r-\La\ .
\ee
where $\rho_m$ and $\rho_r$ corresponds to the energy density associated with $\psi$ and $X$. Now, in an expanding phase the total ``matter'' density $(\rho_r+\rho_m)$ dilutes, and eventually it is canceled by the potential energy, $-\La$, as $H\ra 0$ signalling a ``turnaround'' to a contracting phase~\footnote{Technically, $H\ra0$, does not guarantee a bounce or a turnaround. One could have an eternal universe scenario where $H\ra 0$ as $t\ra0$, or indeed an inflection  point in expansion or contraction. However, from the acceleration equation for the model under consideration, one can easily see that $\ddot{a}<0$ around the point when the cosmological constant cancels the matter contributions, it is therefore a turnaround.}. Thereafter, the matter energy density increases with increasing temperature and eventually when it reaches $\rho_b$, the universe transitions to the expanding branch of the next cycle according to our assumption of the Big Crunch/Bang transition. This then provides us with a cyclic model with a time period of oscillation approximately given by~\cite{emergent-cyclic}~\footnote{In the derivation of the time period it was assumed that the matter density is always dominated by radiation.}.
\be
\cT\approx {\sqrt{3}\pi M_p\over 2 \sqrt{\La}}\,.
\label{time-period}
\ee

The mechanism to obtain inflation in this scenario exploits entropy production when $\psi$ and $X$ interacts. The idea is the following, if $m$ is the mass of $\psi$, then $\psi$ particles are expected to remain in thermal equilibrium with $X$ as long as $T\gtrsim m$ via scattering processes. Below $T=m$,
the massive $\psi$ particles, which are now non-relativistic, will fall out of equilibrium, and consequently
if and when they decay  into radiation  thermal entropy will be generated. This will cause the universe to grow  by making the cycles slightly asymmetric.  After the turnaround, once the temperature becomes sufficiently high, the massive particles are expected to be recreated from radiation via the scattering processes re-establishing thermal equilibrium before the next cycle commences. In order to sustain this growth over many many cyles, what then becomes crucial is to check whether in the contracting universe when the temperature rises above $T=m$, the scattering processes can indeed recreate $\psi$ from $X$ allowing equilibrium to be re-established between NR matter and radiation. While naive arguments by comparing the Hubble rate with scattering rate suggests that this should happen~\cite{cyclic-inflation},  we want to test this assumption first.

Now, the NR particles are expected to decay into massless degrees during the expansion phase prior to the turnaround, let us in fact consider the case when none of these particles are left at the onset of the contraction phase.  To keep the physics simple, we are going to ignore the backreaction of the NR particles on the Hubble expansion rate in this section. It is only going to be relevant for the study of entropy production that we will investigate in the next section.

Let us consider some gauge mediated  creation/annihilation processes between $\psi$ particles and $X$:
\be
\psi+\psi\leftrightarrow X+X\ .
\label{process}
\ee
What we  want to address is really the inverse of the relic/dark matter density computation in conventional Big Bang cosmology. In that  case, we start by assuming that all the species are in thermal equilibrium with each other and ask when the massive species decouples. Here we are going to start by assuming that the $\psi$ particles are not in equilibrium with $X$, and ask whether the scattering processes can recreate them to equilibrium densities.

The Boltzman equation  corresponding to simple processes like (\ref{process}) reads~\cite{kolb-turner}
\be
\dot{n}+3Hn=<\sa |v|>[\nb^2-n^2]\ ,
\label{boltz}
\ee
where $n$ is the number density of the $\psi$ particles, and $\nb$ is the equilibrium number density given by
\be
\nb={1\over 2\pi^2}\int_m^{\infty}{\sqrt{\en^2-m^2}\over e^{\en\over T}\pm 1}\en \ d\en={T^3\over2\pi^2}\int_{m/T}^{\infty}{\sqrt{y^2-(m/T)^2}\over e^{y}\pm 1}y dy\equiv T^3 \bb(T/m)
\label{nb}
\ee
$<\sa |v|>$ is the average cross-section times relative velocity for the process (\ref{process}), and is typically a function of the temperature. If we know  $<\sa |v|>$, in principle (\ref{boltz}) can be solved in conjunction with the Hubble equation and the Boltzman equation for $X$.
\subsection{The NonRelativistic Case}
We will first focus on the nonrelativistic limit, $T<m$, to illustrate the re-equilibriation process as most of the relevant quantities have simple analytic forms in this regime. For the nonrelativistic limit we have
\ba
\nb &=& \LF {mT\over 2\pi}\RF^{3/2}e^{-m/T}\ ,
\label{nr-approx}
\ea
while on general theoretical considerations the thermally averaged  $<\sa |v|>$ is expected to go as a power of $|v|\sim \sqrt{T/m}$. It can thus be parameterized as
\be
<\sa |v|>=\sa_0\LF{T\over m}\RF^n\ .
\label{cross-section}
\ee
For instance, for the $s$ wave channel $n=0$, and for $p$ wave, $n=1$~\cite{kolb-turner}. We will numerically analyze both these cases. $\sa_0$ characterizes the strength of the interaction and would, for instance, be proportional to the square of the appropriate gauge fine structure constant.

According to the usual lore, we expect equilibrium to be established when the interaction rate, $\Ga_s= \nb <\sa |v|>$, catches up with the Hubble expansion rate, $H$, given  by
\be
H^2={\rho_r\over 3M_p^2}={g\pi^2 T^4\over 90 M_p^2}\ .
\label{rad-hubble}
\ee
Above we have used the approximation that at the equilibrium temperature, the universe primarily contains massless particles, $g$  being the number of ``effective'' bosonic  degrees of freedom. Then the condition for re-establishing thermal equilibrium is given by
\be
\Ga_s=\sa_0\LF {mT\over 2\pi}\RF^{3/2}e^{-m\over T}\LF{T\over m}\RF^n\approx \sqrt{g\pi^2\over 90}{T^2\over M_p}
\ee

Defining a new variable $x\equiv T/m$, we can implicitly find the $\xeq$ (or equivalently the temperature) above which equilibrium can be maintained:
\be
{\exp({1\over \xeq}) \sqrt{\xeq}\over \xeq^{n}}=A\al \where A={3\over 2\pi^2}\sqrt{5\over  \pi}\ ,
\label{xeq-estimate}
\ee
and
\be
\al\equiv {\sa_0 m M_p\over \sqrt{g}}
\ee
is the combined dimensionless parameter on which $\xeq$ solely depends.

We are now going to study the process numerically. Substituting   (\ref{cross-section}) in (\ref{boltz}) we have
\be
\dot{n}+3Hn= \sa_0\LF{T\over m}\RF^n\LF\nb^2-n^2\RF\ .
\label{ndot}
\ee
Expressing the time derivative in terms of the temperature derivative, and using (\ref{rad-hubble}) and the fact that $T\propto 1/a$ in radiation dominated universe,  we obtain a reasonably simple equation for the dimensionless variable $\bt\equiv n/T^3$:
\be
{d\bt\over dx}=\al\sqrt{90\over \pi^2}x^n(\bar{\bt}^2-\bt^2)\ ,
\label{beta-x}
\ee
We had previously chosen $\bb$ in such a way that $\bb\equiv \bar{n}/T^3$.

\begin{figure}[htbp]
\begin{center}
\includegraphics[width=0.40\textwidth,angle=0]{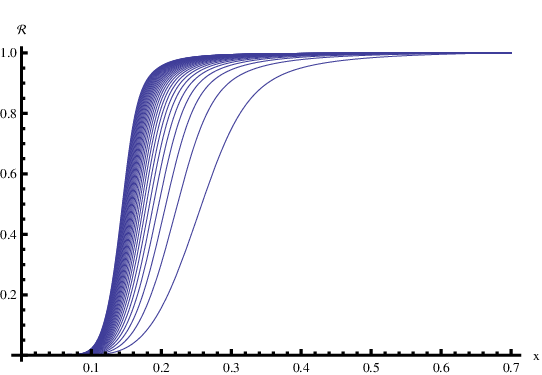}
\includegraphics[width=0.40\textwidth,angle=0]{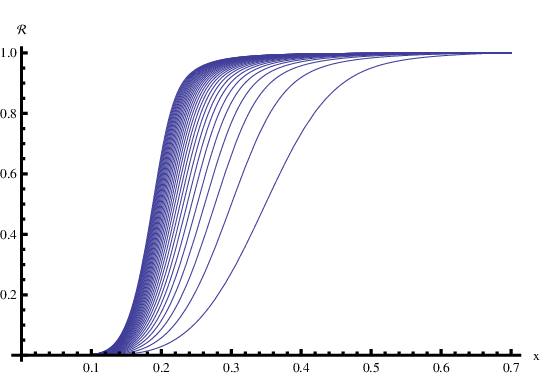}
\end{center}
\caption{{\footnotesize Equilibrium being reached at varying values of $\al$ ranging from 500 to 15000 in steps of 500. Left: $n=0$; Right $n=1$.
\label{fig:NRequilibrium}}}
\end{figure}

In the NR limit we have
\be
 \bar{\bt}(x)={e^{-{1\over x}}\over (2\pi x)^{3/2}}
\ee
To keep track of the departure from equilibrium let us define the ratio
\be
\cR\equiv{n\over \nb}={\bt\over\bb}
\ee
From our numerical analysis, see Fig.~\ref{fig:NRequilibrium}, it is clear that the scattering processes are indeed efficient enough to produce the $\psi$ particles from radiation and establish equilibrium densities. Fig.~\ref{fig:NRequilibrium} also shows that the larger the value of $n$ the longer it takes (higher the temperature) for equilibrium between the NR and relativistic particles to be established; this is to be expected as larger value of $n$ means smaller interaction rate for $x<1$ and hence the longer time-scale needed.

\begin{figure}[htbp]
\begin{center}
\includegraphics[width=0.40\textwidth,angle=0]{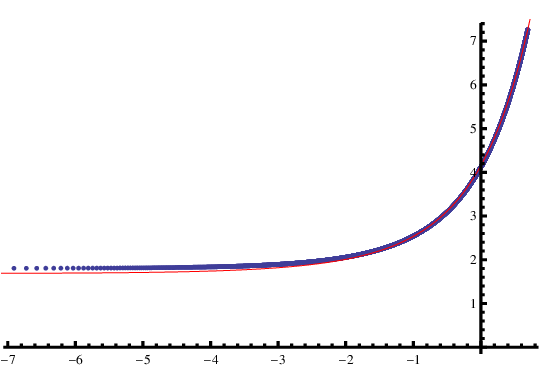}
\includegraphics[width=0.40\textwidth,angle=0]{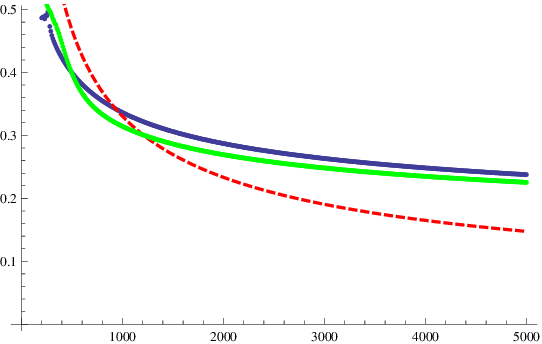}
\end{center}
\caption{{\footnotesize  Left: A comparison between the exact numerical expression of the thermal number density given by (\ref{nb}) and the approximation given by (\ref{rel-approx}). The horizontal axis corresponds to the natural logarithm of $m/T$. We see that almost up to $m/T\sim \cO(10^{-3})$ the approximation is very good. Right: the A plot of $x_{\mt{eq}}=T_{\mt{eq}}/m$ as one changes $\al$. The red dashed line represents the curve based on the naive theoretical expectation given by (\ref{xeq-estimate}),  the blue line corresponds to the numerical value using the Non-relativistic approximation (\ref{nr-approx}) while the green line corresponds to the numerical value using the more accurate approximation formula (\ref{rel-approx}) and $p=0.1$.
\label{fig:x-alpha}}}
\end{figure}

In Fig.~\ref{fig:x-alpha} we have plotted the numerically obtained equilibrium temperature as a function of $\al$ by solving (\ref{beta-x}) in conjunction with the naive estimate given by (\ref{xeq-estimate}). Firstly, it is clear that the more one increases $\al$, the faster one reaches equilibrium \ie, equilibrium is reached at smaller values of $x$ (temperature). This is obviously what one expects and the general trend agrees with the naive estimate (\ref{xeq-estimate}) based on comparison of scattering and Hubble rate. We do note in actuality it takes a little longer to establish thermal equilibrium between $\psi$ and $X$ than  the naive estimate (\ref{xeq-estimate}), but the discrepancy is within the order of magnitude.
\subsection{Relativistic Case}
While there are no phenomenological or theoretical hindrances in looking into the evolution when the $\psi$ particles become relativistic, there are no closed form results for the integral appearing in the equilibrium number density (\ref{nb}) spanning the entire temperature range. Therefore if one were to incorporate it exactly, one would need to solve integro-differential equations which are numerically much more challenging and requires significantly more computational time. Fortunately, we have found an excellent approximation (within a percent) to (\ref{nb}):
\be
\bb(x)\approx e^{\frac{-1}{x}}\left(1.68512+1.94285 x^{-0.84331}-0.72446x^{-1}+\sqrt{\frac{\pi }{2}}x^{-3/2}\right)
\label{rel-approx}
\ee
which can be used for the temperature range $x<10^{3}$ and is sufficient for our purpose~\footnote{In the CI model  the amplitude of the power spectrum is  proportional to $\La^{3/4}/M_p^3$~\cite{BKM-cyclic}. Therefore to obtain the observed value of around $10^{-10}$, one requires $\La^{1/4}\sim 10^{-3}M_p$ and since in our model the variation of the temperature and the mass is limited to $M_p\gtrsim T,m\gtrsim \La^{1/4}$, the range  $x\lesssim 10^{3}$ covers most of the relevant parameter space. }. Fig.~\ref{fig:x-alpha} (left) shows a plot of the exact numerical evaluation vs the approximate analytical function we are using.
\begin{figure}[htbp]
\begin{center}
\includegraphics[width=0.40\textwidth,angle=0]{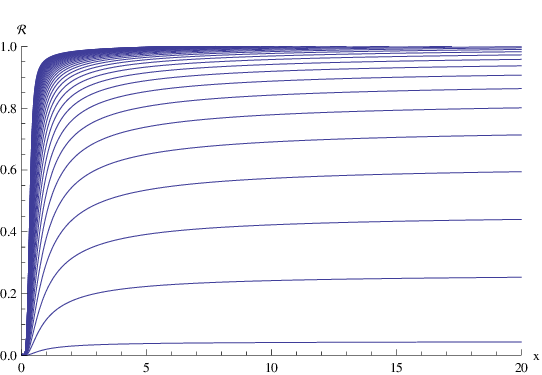}
\includegraphics[width=0.40\textwidth,angle=0]{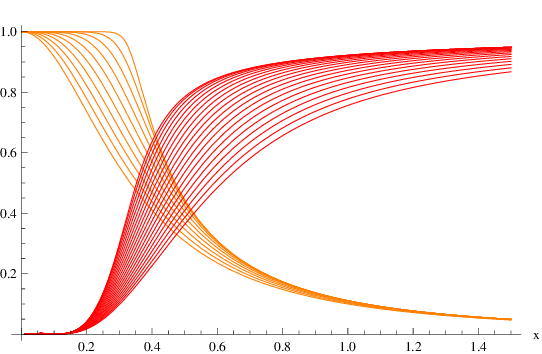}
\end{center}
\caption{{\footnotesize Left: A plot of $x_{\mt{eq}}=T_{\mt{eq}}/m$ with $\al=1\dots 200$ in interval of 5 and $p=0.1$. Right: The orange curve represents plots of the scattering cross-section with $p=0.1\dots 1$ in intervals of 0.1. The smaller the value of $p$, the sharper is the transition from the NR to the relativistic regime. The red curves provides the corresponding plot  for $\cR(x)$  as one changes $p$.
\label{fig:varyingp}}}
\end{figure}

We also need to modify the scattering cross-section (\ref{cross-section}) that we used for the non-relativistic regime. In the relativistic regime, if the interactions are mediated by massless (light) gauge bosons, the cross-section is expected to go as~\cite{kolb-turner}:
\be
<\sa |v|>=\sa_r\LF{m\over T}\RF^2
\label{relativistic-cross}
\ee
where $\sa_r m^2\sim \al_{\mt{f}}^2$, $\al_{\mt{f}}$ being the relevant fine structure constant.  Now, the details of how the scattering crosssection transitions from it's nonrelativistic expression (\ref{cross-section}) to the relativistic one (\ref{relativistic-cross}) depends on the specific high energy physics model in question. Here we will take a phenomenological approach to parameterize the cross-section as
\be
<\sa |v|>={\sa_0\over \LT1+\LF{3x}\RF^{2\over p}\RT^p}\equiv\sa_0 s(x)
\label{model}
\ee
For $x\ll 0.33$ we recover the constant $n=0$ non-relativistic cross-section, while for $x\gg 0.33$, the expression approaches the relativistic approximation with $\sa_r=\sa_0/9$, just a convenient choice to keep the number of parameters tractable. The parameter $p$ controls how fast the transition is from the non-relativistic to the relativistic regime, see Fig.~\ref{fig:varyingp}, right.
The equation we need to solve is then a simple generalization of the non-relativistic version
\be
{d\bt\over dx}=\al\sqrt{90\over \pi^2}s(x)(\bar{\bt}^2-\bt^2)\ ,
\label{Rbeta-x}
\ee
where $\bar{\bt}(x)$ is now given by (\ref{rel-approx}).

Before we go on to study the numerics, let us review some of the well known results pertaining to thermal equilibrium in relativistic species. Since in the relativistic regime (\ref{relativistic-cross}) the interaction rate goes as
$$\Ga_s\sim \bar{n} <\sa |v|>\sim T^3\sa_r\LF{m\over T}\RF^2\ ,$$
equating this with the Hubble parameter one finds that the condition for maintaining thermal equilibrium is given by
\be
{T\over m}\lesssim {\sa_r m M_p\over \sqrt{g}}={\al\over 9}\equiv x_{\max}
\ee
where we have used the simple  relation that exists between $\sa_r$ and $\sa_0$. So at least  in our specific case, the same parameter $\al$ that controls the temperature at which equilibrium is reached, also controls the maximum temperature up to which the equilibrium can be maintained~\footnote{In general if the dominant scattering process responsible for maintaining equilibrium between $\psi$ and $X$ in the relativistic and NR regime, then the parameters controlling the minimum and maximum temperatures for equilibrium will be different.}. What we  found however was that even when the temperatures were way larger than this maximal equilibrium temperature, the $\psi$ density hardly deviates from the equilibrium density, see Fig.~\ref{fig:varyingp} (left). In hindsight, this should not have come as a surprise as it is well known that once a massless species attains equilibrium, the equilibrium densities are automatically maintained even when the excitations are no longer interacting with each other. This is precisely the reason why the CMBR still agrees so well with the Planckian distribution. Thus, it turns out that for the cyclic inflation model, it doesn't make any difference whether the equilibrium is technically lost at some high temperature or not; as long as  equilibrium was established at some temperature in the contracting universe, equilibrium densities are maintained till the beginning of the next cycle. This  adds robustness to the CI mechanism. In general we do not find too much of a difference between the temperatures when equilibrium is reached when we use the NR approximation (\ref{nr-approx}) as opposed to the more exact numerical approximation (\ref{rel-approx})

Our numerical analysis also provides a lower bound for $\al$, below which thermal equilibrium is never attained. For instance, $p=0.1$ at $\al= 70$, one only attains 90\% of the equilibrium density. As we decrease $\al$ further, the maximum ratio $\cR $ attained goes down further. At $\al=0.1$, this ratio is only 0.005. What is interesting  is the ratio remains essentially a constant as the temperature increases towards the bounce, and will essentially be the same for every cycle, see Fig.~\ref{fig:varyingp} (left). Thus the periodicity in energy density of the different species can still be maintained over numerous cycles which is the main ingredient needed for CI mechanism. The CI mechanism appears, clearly, to be more robust than previously thought.

\section{Entropy Production and Growth}~\label{sec:growth}
In the earlier section we saw that, even if we started with no $\psi$ particles, they could be recreated and  thermal equilibrium with radiation could be established as the temperature rose in the contraction phase. What we did not discuss is how the $\psi$ particles are removed from the system in the first place.  The idea, simply, is to make the $\psi$ particles unstable and include decays of $\psi$ into $X$. Such a decay term, in fact, is absolutely crucial for the success of the CI mechanism, as it is responsible for the production of entropy, which manifests itself in the asymmetry of the cyclic evolution leading to a small growth of the scale factor in any given cycle. In this section we want to analyze the entire thermodynamic evolution involving the $\psi$ particles. We want to see how they first fall out of equilibrium during expansion and subsequently decay creating entropy and then finally how they are re-created during contraction. In the process we will also determine whether the universe indeed grows in a given cycle as has been argued, and if so, then how does this growth  depend on  the parameters of the model.

In order to investigate this story we will need to incorporate the decay term, include the backreaction of the NR species on the expansion of the universe, and track $\bt$ as a function of time rather than temperature as we want to include both the expanding and the contracting branches. To keep things tractable we will now specialize to the case when $\psi$ particles always remain non-relativistic, this can be achieved by choosing $m\gtrsim \rhob^{1/4}$, then even at the bounce the temperature is less than $m$.
\subsection{Including the decay term \& Back reaction from matter}
It is relatively straight forward to include the decay term. In the presence of a process such as
\be
\psi\ra X+X
\ee
we need to add a decay term in the continuity equation (\ref{boltz}) for $\psi$:
\be
\dot{n}+3Hn=-\Ga_d n+<\sa |v|>[\nb^2-n^2]\ ,
\label{boltzc}
\ee
where $\Ga_d$, the decay rate is a constant and depends on the details of the specific particle physics processes.  An equation for $\rho_m$ can be obtained straight forwardly by simply multiplying (\ref{boltzc}) by $m$:
\be
\dot{\rho}_m+3H\rho_m=-\Ga_d \rho_m+{\sa_0\over m}\LF{T\over m}\RF^n[\bar{\rho}_m^2-\rho_m^2]\ ,
\label{rho_m}
\ee
Notice, we have switched from using $\bt(t)$ variable to $\rho_m(t)$ for transparency. $\bar{\rho}_m$ is just the equilibrium matter density given by
\be
\bar{\rho}_m=m\bar{n}=mT^3 \bb\LF{ x}\RF
\ee

(\ref{rho_m}) needs to be solved in conjunction with the Hubble equation (\ref{hubble}).
Actually, the Hubble equation is not the best choice to perform numerical calculations involving transitions from expansions and contractions, as one would then need to manually change the sign of the square-root. The way around this technical problem is to use the Raychaudhury equation
\be
{\ddot{a}\over a}=-{1\over 6 M_p^2}\LF\rho_{\tot}+3p_{\tot} \RF=-{1\over 6 M_p^2}\LF {g\pi^2\over 15}T^4+\rho_m+2\La\RF
\label{RC}
\ee
along with the initial conditions given by the Hubble equation.

Finally, we also need an equation for the temperature. This is given by
\be
\dot{T}+HT={15\Ga_d \rho_m\over 2g\pi^2T^3}-{15\sa_0\over 2g\pi^2 T^3m}\LF{T\over m}\RF^n[\bar{\rho}_m^2-\rho_m^2]
\label{T}
\ee
The easiest way to obtain the above is to realize that the total stress energy tensor must be conserved. One can of course obtain the same equation directly by incorporating the interaction terms in the continuity equation for  radiation.

Before solving the equations  (\ref{rho_m}), (\ref{RC}) and (\ref{T}) numerically, let us look at the parameters controlling the physics of the evolution. Naively, there are six parameters, $M_p,m,\sa_0,g,\La$ and $\Ga_d$. However, we know that the asymmetry in the cyclic evolution of the universe is produced as a result of the out-of-equilibrium decay of the massive particles. Such decay processes are only initiated once temperature falls below the equilibrium temperature, and therefore will be governed by $\al$, see the previous section. Clearly, the amount of entropy generated should also depend on the decay rate, or $\Ga_d$, as that controls the entropy production rate. Finally, it should also depend on the amount of time that is available for the decay processes to occur which is dictated by the time period of the cycle, $\cT$ (\ref{time-period}). Thus, on physical grounds we expect the growth in entropy/scale-factor to be dictated only by three independent parameters. Indeed, this becomes manifest if we re-write equations (\ref{rho_m}-\ref{T}) by rescaling some of the variables:
\ba
\rho_m=gm^4R\\
t=\LF{M_p\over \sqrt{g}m^2}\RF\tau
\ea

The above rescaling has a rather simple interpretation; $gm^4$ is approximately the energy density of radiation at the time the massive species becomes non-relativistic. Essentially we are measuring the energy densities in units of this density, and time in units of the time when this transition occurred. In terms of these new variables the equations look
\ba
{a''\over a}&=&-{1\over 6 }\LF {\pi^2\over 15}x^4+R+2\la\RF\\
R'+3R\LF{a'\over a}\RF&=&-\ga R+{g\al}x^n[\bar{R}^2-R^2]\\
x'+x\LF{a'\over a}\RF &=&{15\ga  R\over 2\pi^2x^3}-{15g\al\over 2\pi^2}x^{n-3}[\bar{R}^2-R^2]
\label{coupled-eqns}
\ea
where the set of relevant dimensionless parameters are given by
\ba
\ga&\equiv& {M_p\Ga_d\over \sqrt{g}m^2}\\
\la&\equiv &{\La\over gm^4 }\ ,
\ea
and the combination $g\al$. We have also defined
\be
\bar{R}= {\rho_m\over gm^4}={x^3\over g} \bb\LF{ x}\RF\ ,
\ee
as one can guess.
Again, we are simply measuring $\La$ and $\Ga_d$ in terms of the units associated with the transition of the massive species from relativistic to non-relativistic regime.
\subsection{Factors affecting growth of scale-factor}
\begin{figure}[htbp]
\begin{center}
\includegraphics[width=0.32\textwidth,angle=0]{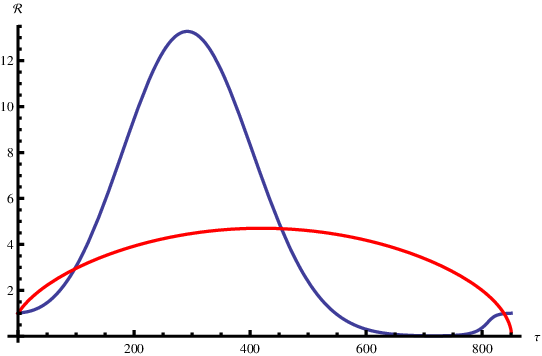}
\includegraphics[width=0.32\textwidth,angle=0]{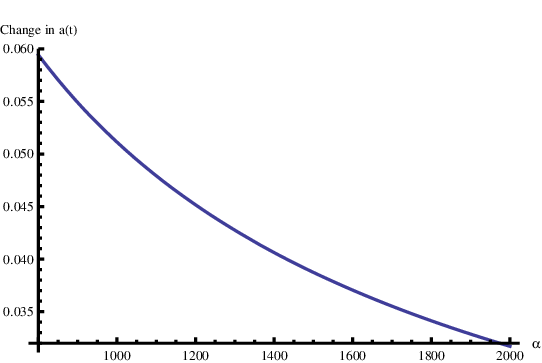}
\includegraphics[width=0.32\textwidth,angle=0]{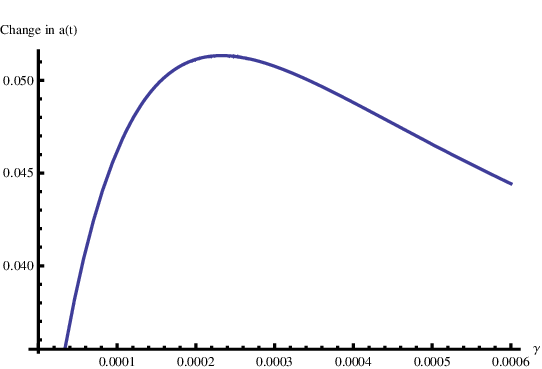}
\end{center}
\caption{{\footnotesize Left: The blue curve tracks $\cR$. It starts out at one and  after the turnaround, again inches back up to one.  The red curve is a plot of the scale factor around the turnaround. The other two figures show how the growth of the scale factor ($\approx \ka$) changes with the parameters of the model. For both the curves we have set $\la=5\times 10^{-8}$. Center: Variation with $\al$, with $\ga=0.0002$ Right: variation with $\ga$, with $\al=1000$.}
\label{fig:dependencies}}
\end{figure}
Since the nonrelativistic limit is only valid as long as $3T<m$, we start our numerical simulations from $x(0)=0.33$. The initial condition for $R$ can be imposed straightforwardly, $R(0)=\bar{R}(x=0.33)$ \ie we start our simulation with equilibrium densities. We can also set $a(0)=1$ as a convention. The initial derivative is then constrained according to the Hubble equation:
\be
\LF{a'\over a}\RF^2=\3\LF R+{\pi^2\over 30}x^4-\la\RF
\ee
Also we choose $g=1$ in all the following simulations.

Fig.~\ref{fig:dependencies} is perhaps the best testament to the viability of the CI scenario. We start with equilibrium density so that the ratio $\cR$ is one. Very quickly, $\psi$ falls out of equilibrium, so it starts to behave like NR matter, diluting more slowly ($\sim a^{-3}$) than it would if thermal equilibrium was maintained (the Boltzman exponential suppression would then be taking over). This is reflected in the $\cR>1$ values. This trend continues till we reach the decay time scale at which point all the $\psi$ particles decay into radiation, generating entropy in the process. Finally, in the contracting phase when the temperature rises again the scattering processes are able to replenish the $\psi$ particles back to equilibrium values. This is exactly what was argued should happen in~\cite{emergent-cyclic,AB-cyclic,cyclic-inflation}, but it is reassuring to verify the mechanism keeping track of all the interactions and back-reactions.

Fig.~\ref{fig:dependencies} shows plots of how the universe grows   during this process as we vary the relevant parameters. The graph at the center exhibits a monotonic decrease in the growth as we increase $\al$. This is easy to understand: as $\al$ increases the universe is able to maintain thermal equilibrium for a longer time and hence the ``out of equilibrium-ness" when the $\psi$ particles do decay becomes less, and  in return less entropy is produced. If $\al$ is so large that $\psi$ is never out of equilibrium, the evolution would be completely symmetric and no entropy would be generated at all.

In Fig.~\ref{fig:dependencies}, the graph on the left shows how the growth is affected as we vary the decay rate and it is a bit more interesting. Initially the growth increases with the increase of the decay rate, and this is obvious. If there is no decay, there is no entropy generation, and as the decay time shrinks the entropy production increases. However, once the decay time scale becomes shorter than the time period, we also start to see a decrease in the growth. This is because, if the $\psi$ particles decay too quickly, they are not sufficiently out of equilibrium when they decay. If in fact the decay time scale becomes smaller than the scattering time scale, we again won't generate any entropy. This is what is represented in the graph.

Before ending this section, let us emphasize that although we used entropy arguments to explain the various features of our numerical results, one could also explain everything in terms of the asymmetry in the NR-matter content between the expanding and contracting phases. In this language the reason why the universe grows a bit more in the expanding branch than it contracts during contraction is because in the former there is a bit more NR-matter than the latter phase, and the universe expands (or contracts) faster in presence of matter than radiation ($t^{2/3}$ as opposed to $\sqrt{t}$).

To summarize, we have verified  through the course of a cycle (really the turnaround) the interactions between the NR and radiative degrees of freedom can produce entropy which causes the universe to expand slightly. We also found the way the growth depends on some of the relevant parameters such as the scattering interaction strength, the decay time and the cycle period, controlled by $\al, \ga, \la$ respectively, were along the expected physical arguments. Although for technical reasons~\footnote{We would not only have to use the approximation (\ref{rel-approx}) for the number density of $\psi$, we would have also had to account for the fact that the number and energy densities are no longer simply related through a constant mass, $m$, but rather through some complex temperature dependent function. Moreover, this would make the equations in terms of energy densities for $\psi$ and $X$ extremely cumbersome, without giving us any new physical insights.} we performed our analysis in the NR regime, we do not expect any qualitative changes if we were to include relativistic regimes. In passing we also note that the optimal value of entropy increase, by around a factor of 0.15  each cycle ($\ka=0.05$), that was found while fitting the WMAP data~\cite{cyclic-prediction} can be easily obtained. For the purpose of illustration, see Fig.~\ref{fig:dependencies} (center and right).
\newpage
\section{Emergent Cyclic Inflation}~\label{sec:ECI}
\subsection{Inflation}
Having seen how the universe can undergo asymmetric cycles it is time to provide a simulation involving many cycles where we can observe the exponential inflationary growth.  In order to achieve this we need to include a last ingredient in our model, a bouncing mechanism. We chose to model the bounce using a $\rho^2$ type correction that has been suggested in Loop Quantum gravity literature~\cite{loop,ashtekar} as well as some models involving brane physics~\cite{Shtanov}. We would like to emphasize, that our primary motivation for choosing this correction is purely technical, it is easy to implement numerically. The details of the bouncing mechanism doesn't affect any of our arguments as they are happening at temperatures much smaller than the bounce temperature, see~\cite{BMS,BGKM,palatini-bounce} for some of the other bouncing mechanisms that has been proposed in the literature. According to the $\rho^2$ bounce the Hubble equation is modified as
\be
H^2={\rho\over 3M_p^2}\LF1-{\rho\over \rho_{\bullet}}\RF
\label{Hbounce}
\ee
As before, for numerical purposes we have to work with the modified Raychaudhury equation:
\be
{\ddot{a}\over a}=-{1\over  M_p^2}\LT\6(\rho+3p)-{\rho\over \rho_{\bullet}}(3p+2\rho)\RT
\label{RCL}
\ee
Here $\rho$ and $p$ refers to the total energy density and pressure:
\ba
\rho&=&gT^4+\rho_m-\La\\
p&=& \3gT^4+\La\ ,
\ea
and $\rho_{\bullet}$ is the energy density at which the universe bounces from a contracting to an  expanding phase as is obvious from (\ref{Hbounce}).

In terms of the rescaled variables the equation reads
\be
{a''\over a}=-\LT{1\over 6 }\LF {\pi^2\over 15}x^4+R+2\la\RF-\ka\LF {\pi^2\over 30}x^4+R-\la\RF\LF {\pi^2\over 10}x^4+2R+\la\RF\RT
\ee
where
\be
\ka\equiv {gm^4\over \rho_{\bullet}}
\ee
The appropriate initial condition is given by
\be
\LF{a'\over a}\RF^2=\3\LF R+{\pi^2\over 30}x^4-\la\RF\LT1-\ka\LF R+{\pi^2\over 30}x^4-\la\RF\RT
\label{mod-hubble}
\ee

\begin{figure}[htbp]
\begin{center}
\includegraphics[width=0.30\textwidth,angle=0]{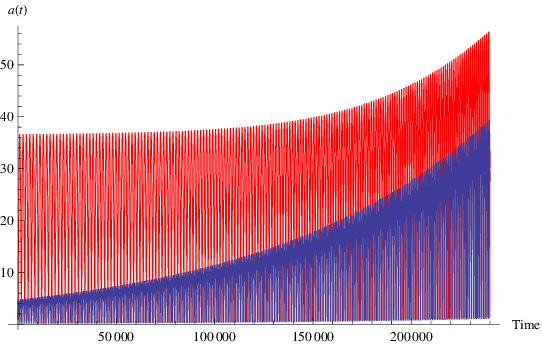}
\includegraphics[width=0.30\textwidth,angle=0]{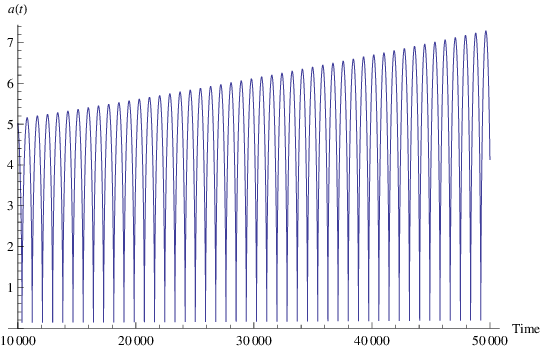}
\includegraphics[width=0.35\textwidth,angle=0]{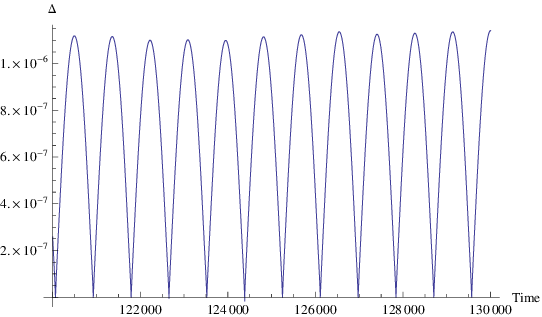}
\end{center}
\caption{{\footnotesize Left: The evolution in blue corresponds to the cyclic inflationary growth while the plot in red corresponds to the evolution for a closed universe. It is clear that the universe in the latter case asymptotes to a periodic evolution in the past.  Center: We have zoomed in to see how the growth occurs over a few consecutive cycles in the cyclic inflationary phase. Right: We have plotted $\Da$ which measures the accuracy to which the first integral of motion, (\ref{Hbounce}), is satisfied.}
\label{fig:exponential-growth}}
\end{figure}

Fig.~\ref{fig:exponential-growth} shows the evolution of the scale factor over numerous cycles. The evolution is nonsingular. Although the bounces represent a sharp transition from contraction to expansion, it is driven by the $\rho^2$ modification in (\ref{Hbounce}) known to produce nonsingular bounces. Evidently, the evolution shows an exponential growth over many cycles. Indeed the cyclic inflation mechanism works!

We simulated our differential equations numerically using Mathematica 9 which allowed us to ensure the accuracy of our simulations to a high degree by appropriately increasing the ``accuracy goals''~\cite{mathematica-accuracy} that controls ``effective digits of accuracy'' for the various numerical operations within the Mathematica code. To account for the ``stiff'' bounce phase where the scale factor changes rather quickly, we adopted the ``StiffnessSwitching'' technique to numerically solve the differential equations which  switches between a ``nonstiff and a stiff solver'' method to preserve the accuracy and efficiency of the programming. For the nonstiff solver we used the ``explicitly modified midpoint'' method while for the stiff solver we used the ``linearly implicit Euler'' method~\cite{mathematica}. We also made sure that the maximum time step for the numerical simulations is much smaller than the bounce time scale, the shortest time scale in the physical problem. Since we used the acceleration equation to obtain our evolution, there is an important consistency check we can perform for our numerical analysis. The modified Hubble equation (\ref{Hbounce}) should remain valid throughout the evolution. We have verified this at the precision level of $10^{-6}$ or better.  Please see Fig.~\ref{fig:exponential-growth} where we have plotted $\Da \equiv (LHS-RHS)/T^4$, where $LHS,RHS$ correspond to the left and right hand side of (\ref{Hbounce}) respectively, and $T^4$ represents the typical energy density scale. This is particularly crucial because it also demonstrates that the numerical errors did not accumulate through the integration process over many cycles.

Finally, because of the data-size and rapidly changing quantities of our solutions one can sometimes have trouble effectively visualizing these data plots. This can be circumvented by increasing the recursion limits of the plotting techniques in Mathematica which can be set to reiterate a plotting algorithm up to 15 times in an attempt to maintain resolution at a cost of speed.

Coming back to physics, we would like to re-emphasize that this cyclic behavior needs to come to an end and
usher in a stage of monotonic expansion. This is indeed possible if one upgrades the cosmological constant to the potential energy coming from various scalar (moduli) fields, the scalar field can then roll along a relatively flat negative potential energy region before making a classical jump to a positive energy region within a single contracting phase. This happens for quite natural initial conditions and have been studied in~\cite{BKM-exit}.  We therefore do not elaborate on this mechanism here.
\subsection{Emergence}
Let us finally revisit the issue of geodesic completeness in the context of the CI scenario. One of the main theoretical benefits of the CI model as compared to the traditional inflationary scenario is that it is nonsingular and can be extended into the infinite past. But is that really true? If one tracks, say, the maxima/turnarounds of the oscillating space time, then one finds that it has the traditional inflationary trajectory, so the problem of past geodesic incompleteness apparently comes back to haunt us! Fortunately, there appears a natural resolution for a closed universe model which was first put forward in~\cite{emergent-cyclic} in a slightly different setting: As one goes back in cycles, there comes a point when the curvature energy density becomes more important than the vacuum energy density. (Curvature density blue shifts as $a^{-2}$, while the vacuum energy density remains a constant.) Once this happens, the universe no longer turns around due to the negative vacuum energy density, but before, when $\rho_r+\rho_k=0$. Let us then look at the dependency of the turnaround temperature on the entropy $S$ of the universe. A convenient choice for the scale factor and the metric is given by~\cite{misner-thorne-wheeler}
\be
ds^2=-dt^2+a^2(t)\LF{dr^2\over 1-kg m^4r^2/3M_p^2}+r^2d\Oa^2\RF\mand a =g^{\4}mV^{1/3}
\label{convention}
\ee
$V$ being the actual volume of the universe.
According to this convention the ``energy density'' associated with  radiation and curvature in terms of the scale factor is given by~\footnote{We have used  elementary thermodynamic relations, such as $S={4\rho_rV\over 3T}$ and $\rho_r={g\pi^2 T^4\over 30 }$ as well as the definition of the scale factor (\ref{convention}) to obtain $\rho_r$ as a function of the scale factor and entropy.}
\ba
\rho_r=Bgm^4{S^{4/3}\over a^4}\mand \rho_k= -gm^4{k\over a^2}\with B\equiv \LF{3\over 4}\RF^{4/3}\LF{30\over g\pi^2}\RF^{1/3}
\ea
$S$ represents the entropy of the universe for the given cycle, so it keeps on decreasing as we go back in cycles. The turnaround occurs when $\rho_r$ and $\rho_k$ cancel each other, so that
\be
a_{\mt{turn}}={\sqrt{B}S^{2/3}\over \sqrt{k}}
\ee
Further, since
\be
a\propto {S^{1/3}\over T}
\ee
we have
\be
T_{\mt{turn}}\propto {\sqrt{k}\over S^{1/3}}
\ee

In other words the smaller the entropy, the quicker the turnaround, and the higher the temperature of the turnaround. In turn, as the turnaround temperature becomes closer to the temperature when equilibrium is lost, the smaller is the entropy produced, and therefore the smaller is the growth of the universe in a given cycle. Thus the universe no longer grows by the same factor in every cycle but the growth starts to become vanishingly small as we go back in cycles, see~\cite{emergent-cyclic} for a more detailed analytical argument. Effectively, as one goes back in cycles the entropy production switches off~\footnote{For a universe where time can be continued till past infinity, it becomes a bit confusing to think in terms of “initial conditions”. Rather, for instance, one can think about the entropy of the present universe. As was first discussed in~\cite{emergent-cyclic}, there is a critical value of entropy below which  the evolution of the universe is periodic. In this case, there is no entropy production in any of the cycles as the turnaround temperature is higher than the temperature when equilibrium is lost. If, on the other hand, the entropy of the universe today is larger than this critical value, this implies an emergent cyclic behavioral pattern where  the entropy has been increasing in every cycle. The increase however tapers down as $t\ra -\infty$, and the entropy approaches the critical value from above as $t \ra -\infty$. In this case the entropy production never really completely switches off, but approximately does so. Mathematically, when one solves the coupled set of differential equations governing the evolution of the scale factor and the energy components, two different sets of trajectories emerge: Periodic trajectories which are phenomenologically unviable, and the asymmetric emergent cyclic trajectories which can be viable. } and the universe asymptotes to a periodic evolution which is indeed geodesically complete.

To see this play out quantitatively, we have to include a curvature term in the energy density and pressure:
\ba
\rho&=&\rho_r+\rho_m-\La+\rho_k\\
p&=& \3\rho_r+\La-\3\rho_k\ ,
\ea

For our numerical simulation, all the equations stay the same except the Hubble and the RC equations which now read
\ba
\LF{a'\over a}\RF^2&=&\3\LF R+{\pi^2\over 30}x^4-\la-{k\over a^2}\RF\LT1-\ka\LF R+{\pi^2\over 30}x^4-\la-{k\over a^2}\RF\RT\label{Hbounce-curv}\\
{a''\over a}=&-&\LT{1\over 6 }\LF {\pi^2\over 15}x^4+R+2\la-{k\over a^2}\RF\Rd\non
&-&\Ld\ka\LF {\pi^2\over 30}x^4+R-\la-{k\over a^2}\RF\LF {\pi^2\over 10}x^4+2R+\la-2{k\over a^2}\RF\RT\label{accel-curv}
\ea
As before, we perform our numerical calculations with the acceleration equation (\ref{accel-curv}). The numerical results are illustrated in Fig.~\ref{fig:exponential-growth}. One clearly sees that the addition of the curvature term ameliorates the growth of the scale factor as we go back in cycles; the space-time is asymptoting to a periodic evolution. Again, we have checked that the first integral of motion, (\ref{Hbounce-curv}), is satisfied. This is a first numerical evidence of the emergence mechanism and it is certainly good news that this mechanism can smoothly transition into the cyclic inflation phase.
\section{Summary and Future Directions}\label{sec:conclusions}
We have demonstrated in this paper a viable emergent cyclic model of inflation which
relies on two major assumptions:  (i) There exists a robust mechanism that can induce a
smooth bounce in the early universe as and when the energy density approaches Planckian values. Our
model only relies on the existence of a bounce mechanism; the details of the mechanism are not really important, we used a loop quantum cosmology motivated bounce for technical simplicity,  but one is free to envision other avenues. (ii) The cosmological constant is not a constant but instead should be thought of as a scalar field(s) dependent potential that has the ability to shift between
positive and negative values. Although this aspect did not enter our investigations as we were primarily interested in the emergent and inflationary regime, this is relevant to provide a graceful exit from the CI phase. The reader is referred to~\cite{BKM-exit} for details, also see~\cite{Piao:2004me}. The only other assumption of our model, namely that in the ``beginning'' ($t\ra -\infty$) the patch of universe under consideration contained both massive and massless excitations interacting with each other via scattering and decay processes, can hardly be called an assumption, as it is  what one normally expects.

We expect the CI to be able to address the usual cosmological puzzles, such as the ones associated with the problems of
isotropy (or Mixmaster behavior), horizon, flatness and homogeneity (or Black hole over-production) in our universe~\cite{cyclic-inflation}, in a manner very similar to the standard inflation. For instance, since
anisotropies $\propto a^{-6}$, once the cyclic-inflationary phase is ``activated'' in a small and
sufficiently smooth patch of the universe the chaotic Mixmaster behavior is avoided in
subsequent cycles because the scale factor at the consecutive bounce points also keep
growing and the universe becomes more and more isotropic. Very similar reasoning
also resolves the flatness problem; in fact, this is the reason we see a transition from the emergent (curvature mediated bounce) to the inflationary (vacuum energy mediated bounce) phase. The issue about the growth of inhomogeneities is more subtle and we are presently investigating it in more details, but here are some general arguments which lead us to believe that the CI scenario is safe from the problem of inhomogeneities/black hole overproduction. Firstly, let us point out that small scale inhomogeneities (subHubble) can only grow above the Jeans length. In the CI scenario, radiation is always the dominant matter component (we only need small amounts of nonrelativistic species to implement CI, and in fact, they all decay during the expansion phase in each cycle) and hence the Jeans length is comparable to the Hubble length. In other words, the radiation pressure ensures that there is no small scale growth of density fluctuations. On the other hand, since the universe undergoes an overall inflationary growth, we expect superHubble fluctuations to remain a constant as in standard inflation. In fact, one can calculate the primordial spectrum of fluctuations by computing the Bardeen potential at the time of Hubble crossing. This was first discussed in~\cite{cyclic-prediction} where it was found that the cycic inflation mechanism produces oscillatory features on top of the usual near-scale-invariant scalar spectrum. Imprints of thermal fluctuations was calculated in~\cite{BKM-cyclic}.  In~\cite{BKM-hagedorn} another rather interesting possibility of producing a relatively large tensor-to-scalar ratio and a blue gravity wave tilt was proposed that can easily be applied to the CI scenario. This would however require extending the model investigated in this paper with stringy Hagedorn phases~\cite{Atick-witten}. One would also need to study the propagation of the perturbations around the bounce, as bounces may have significant effects as shown in~\cite{gw-cyclic}. We leave these exercises for future.

The major advantages of this model include (i) providing a non-singular and
 geodesic complete evolution, (ii) providing the opportunity to explore an early universe that
contains a negative value of potential energy (the existence of which is strongly suggested by String/Supergravity theory), (iii)  obviating the need for any reheating mechanism at the end of inflation, as radiation is always the dominant energy density component. While we consider these the primary motivations for such a model it is up to the reader to imagine how
new and exciting ideas such a scenario may be able to facilitate. For instance, an aspect worth exploring in the context of the CI paradigm is the production of particle/anti-particle asymmetry. While many different options exist (see for instance~\cite{Krylov:2013qe,Shaginyan:2011zz,Volkas:2013za}) that can produce small violations of matter/anti matter asymmetry in the standard cosmological model it still is challenging to find a mechanism that can generate the large asymmetry needed to be consistent with observations. This is largely due to the fact that such an asymmetry can be developed only, in
most mechanisms, during particular phase transitions of the early universe. According to the popular inflationary
model, the universe spends only a brief amount of time in these regimes making it difficult to produce the desired violation of particle/anti-particle symmetry. One of the virtues of the ECI model is that innumerable such windows exist! At each
cycle some amount of asymmetry can be produced (see~\cite{littlebangs} for a relevant discussion) as the universe shifts in and out of
equilibrium.

Another such idea, could be to try to connect the process of entropy production, so central to the ECI paradigm, to gravity itself. In~\cite{Verlinde}  Verlinde proposed gravity as a statistical phenomenon derived from an ``entropic force''. Curiously, the entropic force, which tries to restore the system to an equilibrium state, is shown to be proportional to the temperature. Therefore in a given cycle we would expect this force to succeed in attaining equilibrium near bounces, but then as the universe expands and the temperature falls, out of equilibrium phases could possibly prevail, and this pattern of thermodynamic events is the essential ingredient that is needed for the success of the ECI mechanism!  This would certainly explain why entropy is so important in the early universe?

To conclude, the current paper provides a comprehensive numerical analysis essentially validating the ECI scenario. However, the model will  be far more interesting if there are any observational evidences. The CI model, in fact,  predicts rather distinctive signatures in the form of characteristic wiggles on top of the near scale-invariant spectrum. A preliminary fit with the WMAP data yielded interesting results, and thus it is imperative that we carry out a comprehensive data-fitting exercise with the newly released Planck data. This is where our current analysis would be useful as the amplitude and period of the oscillatory features  depend on the factor by which the universe grows in a single cycle, and we now understand how this can be connected with model parameters such as $\al,\ga, \la$. We would also need to comprehensively analyze and include the evolution of the scalar field modulated potential energy (replacing the cosmological constant) to especially determine how the amplitude of fluctuations is changing with the wave number (spectral tilt), and this we also leave for future investigations.
\vs
{\bf Acknowledgments:}
WD would like to specially thank  Mr. Bob Hanlon for his guidance throughout the development of the simulations. TB's research has been supported by the LEQSF(2011-13)-RD-A21 grant from the Louisiana Board of Regents.
\bibliography{cyclicrefs}
\bibliographystyle{ieeetr}
\end{document}